\documentclass[twocolumn,showkeys,aps,prb,showpacs]{revtex4-1}
\usepackage{graphicx}
\usepackage[CJKbookmarks,dvipdfm,colorlinks,linkcolor=blue,citecolor=blue]{hyperref}

\begin{document}

\title{Potential 2D thermoelectric materials ATeI (A=Sb and Bi) monolayers from a first-principles study}

\author{San-Dong Guo and Ai-Xia Zhang}
\affiliation{$^1$School of Physics, China University of Mining and
Technology, Xuzhou 221116, Jiangsu, China}

\begin{abstract}
Lots of two-dimensional (2D) materials have been predicted  theoretically,  and further confirmed  in experiment, which have wide applications in nanoscale electronic, optoelectronic and  thermoelectric devices.  Here, the thermoelectric properties of ATeI (A=Sb and Bi) monolayers are  systematically investigated, based on semiclassical Boltzmann transport theory.
It is found that spin-orbit coupling (SOC)  has important effects on  electronic   transport coefficients in p-type doping, but neglectful influences on n-type ones.  The room-temperature sheet thermal conductance  is 14.2 $\mathrm{W  K^{-1}}$ for SbTeI and 12.6 $\mathrm{W  K^{-1}}$ for BiTeI,  which are lower than one of most well-known 2D materials, such as transition-metal dichalcogenide, group IV-VI, group-VA and group-IV monolayers. By  analyzing group  velocities and  phonon lifetimes, the very low sheet thermal conductance of ATeI (A=Sb and Bi) monolayers is mainly due to small group  velocities.
It is found  that the high-frequency optical branches contribute significantly to the total thermal conductivity,  being obviously different from usual picture with little contribution from optical branches. According to cumulative lattice thermal conductivity   with respect to phonon mean free path (MFP), it is difficulty to further reduce lattice thermal  conductivity by nanostructures.
Finally,  possible thermoelectric figure of merit $ZT$ of  ATeI (A=Sb and Bi) monolayers are calculated. It is found that  the p-type doping has more excellent thermoelectric properties than n-type doping, and at room temperature,  the peak $ZT$ can reach 1.11 for SbTeI and 0.87 for BiTeI, respectively.  These results  make us believe that ATeI (A=Sb and Bi) monolayers may be potential 2D thermoelectric materials, and can  stimulate further experimental works to synthesize  these  monolayers.

\end{abstract}
\keywords{Monolayer; Lattice thermal conductivity; Group  velocities; Phonon lifetimes}

\pacs{72.15.Jf, 71.20.-b, 71.70.Ej, 79.10.-n ~~~~~~~~~~~~~~~~~~~~~~~~~~~~~~~~~~~Email:guosd@cumt.edu.cn}

\maketitle

\section{Introduction}
Due to their potential  applications in energy-related issues, thermoelectric materials have been widely investigated both in experiment and theory\cite{q0,q1}. The  performance  of a thermoelectric
material is measured by the dimensionless figure of merit $ZT$, defined as $ZT=S^2\sigma T/(\kappa_e+\kappa_L)$, in which S, $\sigma$, T, $\kappa_e$ and $\kappa_L$ are the Seebeck coefficient, electrical conductivity, working temperature, the electronic and lattice thermal conductivities, respectively. To attain a high $ZT$ value, a high power factor ($S^2\sigma$) and/or a low thermal conductivity ($\kappa=\kappa_e+\kappa_L$) are required. Unfortunately, these transport coefficients of  bulk materials are coupled with each other, which are oppositely  proportional to carrier density. Therefore, searching for high-performance thermoelectric materials is challenging.

The $ZT$ values of thermoelectric materials can be significantly enhanced  by using low-dimensional
systems or nanostructures, which is firstly proposed by Hicks and Dresselhaus in 1993\cite{q2,q3}.
A large number of subsequent works have been centered on nanostructured materials\cite{q4,q5,q6},  questing for highly efficient thermoelectric materials.  Lots of 2D monolayers have been predicted in theory, or have been synthesized experimentally, including semiconducting transition-metal dichalcogenide\cite{q7} (such as $\mathrm{MoS_2}$ and $\mathrm{PtSe_2}$), group IV-VI\cite{q8} (such as SnS and SnSe), group-VA\cite{q9,q10} (such as arsenene and antimonene) and group-IV\cite{q11} (such as germanene and stanene) monolayers. To design high-performance
thermoelectric devices, the thermoelectric properties related with these 2D materials have also been investigated.
\begin{figure}
  \includegraphics[width=5.2cm]{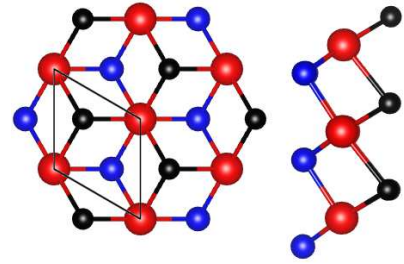}
  \caption{(Color online) Top and side view of the crystal structure
of ATeI (A=Sb and Bi) monolayers. The large red balls represent A atoms, and the small blue balls for Te atoms, and the smallest black balls for I atoms. The frame surrounded by a black box is  unit cell.}\label{st}
\end{figure}
  Based on ab-initio method and ballistic transport model,the thermoelectric properties of $\mathrm{MX_2}$ (M=Mo, W; X=S, Se) monolayers have been studied\cite{q12}, and  a maximum $ZT$ of monolayer $\mathrm{MoS_2}$  is up to  0.5 at room temperature. For monolayer $\mathrm{MoS_2}$, a value of S as 30 mV/K  has been reported experimentlly\cite{q13}.
It has been proved that strain  is a very  effective strategy to improve  thermoelectric properties of monolayer $\mathrm{PtSe_2}$ by enhancing power factor and reducing lattice thermal conductivity\cite{q14}.
The transport coefficients of  orthorhombic group IV-VI monolayers $\mathrm{AB}$ (A=Ge and Sn; B=S and Se)
 have been  systematically investigated  theoretically\cite{q15,q16}. The  lattice  thermal conductivities of  $\alpha$- and $\beta$-As, Sb monolayers  have been performed in theory\cite{q17,q18,q19,q20}.  The  thermoelectric properties of $\alpha$-As monolayer have been investigated with Green's function based transport techniques\cite{21}, and  thermoelectric properties of $\beta$-Bi monolayer have also been studied by equilibrium molecular dynamics simulations\cite{q22}.
Phonon transport properties of  group-IV monolayers (graphene, silicene, germanene and stanene) have been systematically investigated  from ab initio calculations\cite{q23}, and it is found that the lattice
thermal conductivity  for graphene, silicene and germanene decreases monotonically, but  higher  lattice
thermal conductivity is observed in stanene.

\begin{figure}
  \includegraphics[width=8cm]{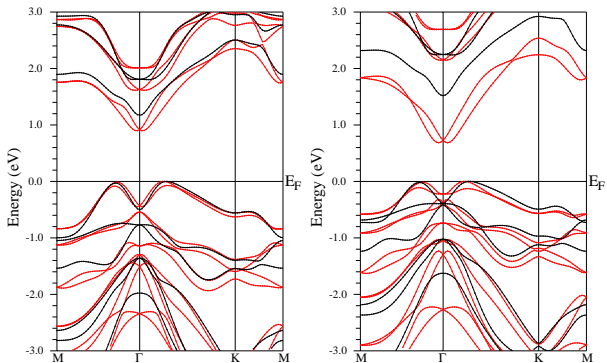}
  \caption{(Color online) The energy band structures of SbTeI (Left) and BiTeI (Right)  using GGA (Black lines) and GGA+SOC (Red lines).}\label{band}
\end{figure}
\begin{figure}[!htb]
  \includegraphics[width=7cm]{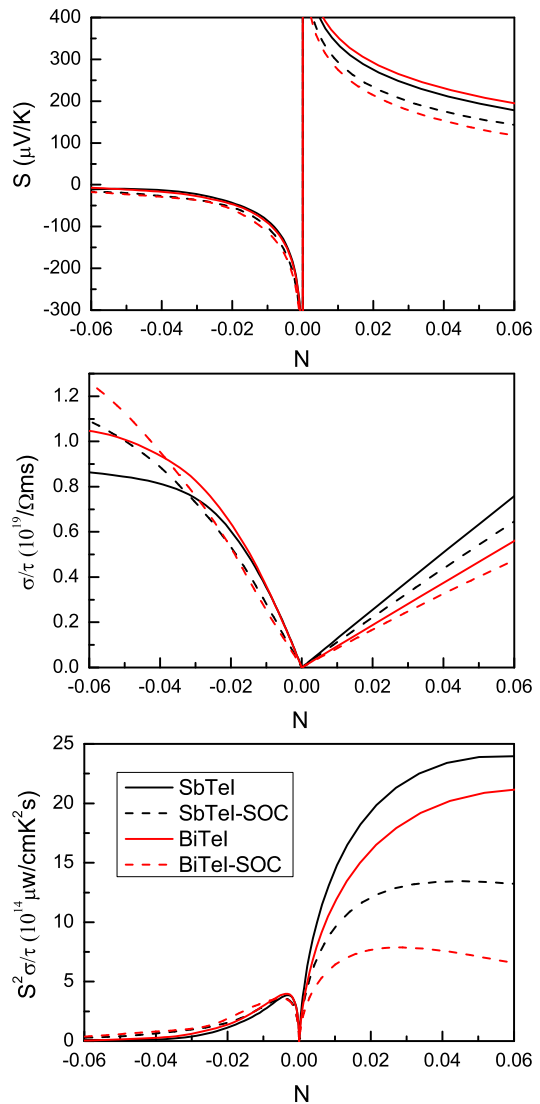}
  \caption{(Color online) At room temperature (300 K),   Seebeck coefficient S,  electrical conductivity with respect to scattering time  $\mathrm{\sigma/\tau}$ and  power factor with respect to scattering time $\mathrm{S^2\sigma/\tau}$ of SbTeI and BiTeI   using GGA  and GGA+SOC as a function of doping level (N).}\label{s}
\end{figure}

Recently, based on the first-principles calculations, stable ATeI (A=Sb and Bi) monolayers  have been predicted with a giant Rashba spin splitting\cite{q24,q25}. The thermoelectric properties related with bulk BiTeI have been performed\cite{q26}, and the figure of merit $ZT$ at 300 K is 0.05, and the $ZT$ of undoped BiTeI reaches 0.3 at 520 K. The  thermoelectric
performance of bulk BiTeI  can be improved in Cu-intercalated BiTeI \cite{q27} or through Br-substitution\cite{q28}. The pressure-enhanced power factor has been predicted by the first-principles calculations in bulk BiTeI\cite{q29}.
Here, we systematically study  the thermoelectric performance of ATeI (A=Sb and Bi) monolayers by  combining the first-principles calculations and semiclassical Boltzmann transport theory. It is necessary for calculations of electronic   transport coefficients of ATeI (A=Sb and Bi) monolayers to include SOC.  The calculated sheet thermal conductances of ATeI (A=Sb and Bi) monolayers are lower than ones of other well-studied 2D materials.
The contribution to total thermal conductivity
from high-frequency optical branches is
larger than 19\%. This is   different from the usual picture  that high frequency
optical branches have  very little contribution. Based on calculated $ZT$, ATeI (A=Sb and Bi) monolayers  may be potential 2D
thermoelectric materials.

The rest of the paper is organized as follows. In the next
section, we shall give our computational details about calculations of electronic structures, electron and phonon transport. In the third section, we shall present electronic structures, electron and  phonon transport of ATeI (A=Sb and Bi) monolayers. Finally, we shall give our discussions and conclusions in the fourth section.

\section{Computational detail}
A full-potential linearized augmented-plane-waves method
within the density functional theory (DFT) \cite{1} is employed to carry out electronic structures of  ATeI (A=Sb and Bi) monolayers, as implemented in the  WIEN2k code \cite{2}. The GGA of Perdew, Burke and  Ernzerhof  (GGA-PBE)\cite{pbe} is used to  optimize free  atomic position parameters   with a force standard of 2 mRy/a.u..
The SOC is included self-consistently \cite{10,11,12,so} due to large Rashba
spin splitting, which produces important effects on electronic transport coefficients. The convergence results are determined
by using  4000 k-points in the
first Brillouin zone (BZ) for the self-consistent calculation, making harmonic expansion up to $\mathrm{l_{max} =10}$ in each of the atomic spheres, and setting $\mathrm{R_{mt}*k_{max} = 8}$ for the plane-wave cut-off. The self-consistent calculations are
considered to be converged when the integration of the absolute
charge-density difference between the input and output electron
density is less than $0.0001|e|$ per formula unit, where $e$ is
the electron charge.

Based on calculated energy band
structures, transport coefficients of electron part, including Seebeck coefficient and electrical conductivity,
are calculated through solving Boltzmann
transport equations within the constant
scattering time approximation (CSTA),  as implemented in
BoltzTrap code\cite{b}, which shows reliable results in many classic thermoelectric
materials\cite{b1-1,b2,b3}. To
obtain accurate transport coefficients,  the parameter LPFAC  is set as 20, and  at least 2000 k-points is used in the  irreducible BZ for the calculation of energy band structures. The  lattice thermal conductivities are performed
by using Phono3py+VASP codes\cite{pv1,pv2,pv3,pv4}. The second order harmonic and third
order anharmonic interatomic force constants (IFC) are calculated by using a  5 $\times$ 5 $\times$ 1   supercell  and a  3 $\times$ 3 $\times$ 1 supercell, respectively. To compute lattice thermal conductivities, the
reciprocal spaces of the primitive cells  are sampled using the 40 $\times$ 40 $\times$ 2  meshes.

For 2D material, the calculated  electrical conductivity, electronic  and lattice  thermal conductivities  depend on the length of unit cell along z direction\cite{2dl}.  They  should be normalized by multiplying Lz/d, where Lz is the length of unit cell used in the calculations along z direction  and d is the thickness of 2D material, but the d  is not well defined.  However, the dimensionless figure of merit $ZT$ is independent of  the length of unit cell along z direction. In this work, the length of unit cell (20 $\mathrm{{\AA}}$)  along z direction is used as the thickness of  ATeI (A=Sb and Bi) monolayers. By $\kappa$ $\times$ d,  the thermal sheet conductance can be attained, which is  used to compare the  thermal conductivities of various  2D monolayers.

 \begin{table}[!htb]
\centering \caption{The  lattice constants\cite{q25} $a$ ($\mathrm{{\AA}}$); the calculated energy band gaps  using GGA $G$ (eV) and GGA+SOC $G_{so}$ (eV); $G$-$G_{so}$ (eV);  Rashba energy $E_{R}$ (meV). }\label{tab}
  \begin{tabular*}{0.48\textwidth}{@{\extracolsep{\fill}}ccccccc}
  \hline\hline
Name& $a$ &  $G$& $G_{so}$&$G$-$G_{so}$& $E_{R}$\\\hline\hline
SbTeI&4.32&1.17&0.90&0.27& 18\\\hline
BiTeI&4.42&1.52&0.69&0.83&42\\\hline\hline
\end{tabular*}
\end{table}

\begin{figure}
  \includegraphics[width=7cm]{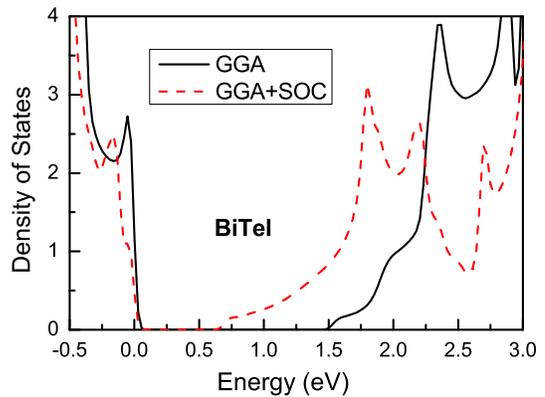}
  \caption{(Color online) The DOS of  BiTeI   using GGA  and GGA+SOC.}\label{dos}
\end{figure}

\begin{figure}
  \includegraphics[width=8cm]{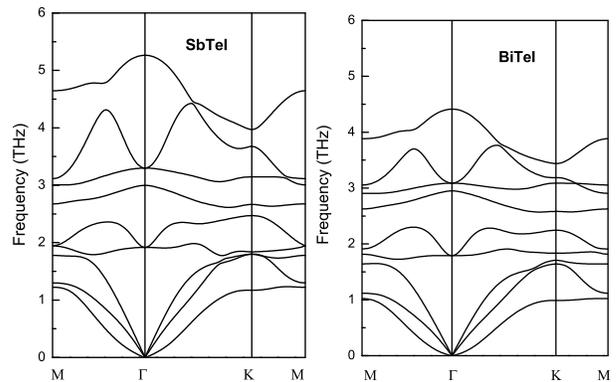}
  \caption{Phonon band structure of SbTeI (Left) and BiTeI (Right) monolayers using GGA-PBE.}\label{phband}
\end{figure}

\section{MAIN CALCULATED RESULTS AND ANALYSIS}
ATeI (A=Sb, I) monolayers  can be attained from their bulk
counterparts  with a trigonal structure, and they  consist of three sublayers with A atoms in the center sublayer, while  Te and I atoms  are located in the top and bottom sublayers.  The schematic crystal structure is plotted  in \autoref{st}. The unit cell  of  ATeI (A=Sb, I) monolayers, containing one A, one  Te and one I atoms,  is built with the vacuum region of larger than 15 $\mathrm{{\AA}}$ to avoid spurious interaction. The optimized lattice constants\cite{q25} within GGA-PBE are used to investigate their electronic structures and thermoelectric properties, and free atomic positions are also optimized. \autoref{band}  shows the energy band structures of ATeI (A=Sb, I) monolayers using both GGA and GGA+SOC. At the absence  of SOC,   ATeI (A=Sb, I) monolayers are both semiconductors with the
conduction band minimum (CBM) at the $\Gamma$ point, and with two valence band maxima (VBM)  between the $\Gamma$ and K or M points. When the SOC is considered, two CBM are located
slightly shifted away from $\Gamma$ point along the $\Gamma$-K and $\Gamma$-M paths due to Rashba effect, and the energy band gap is reduced from 1.17 eV to 0.90 eV for SbTeI, and from 1.52 eV to 0.69 eV for BiTeI. The Rashba energy of ATeI (A=Sb, I) monolayers, defined as the energy difference between the CBM and the band crossing point of conduction bands at  $\Gamma$ point,
 is 18 meV and 42 meV, respectively. Some key data are shown in \autoref{tab}, which agree well with previous results\cite{q25}.

\begin{figure}
  \includegraphics[width=6cm]{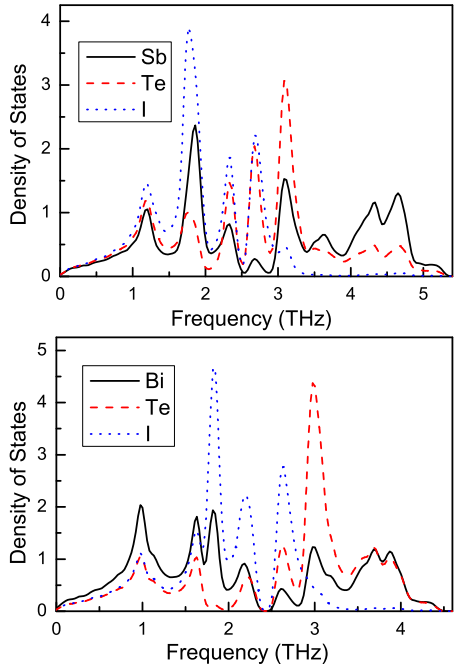}
  \caption{Phonon partial  DOS of  SbTeI (Top) and BiTeI (Bottom) monolayers  using GGA-PBE.}\label{ftdos}
\end{figure}
\begin{figure}
  \includegraphics[width=7.0cm]{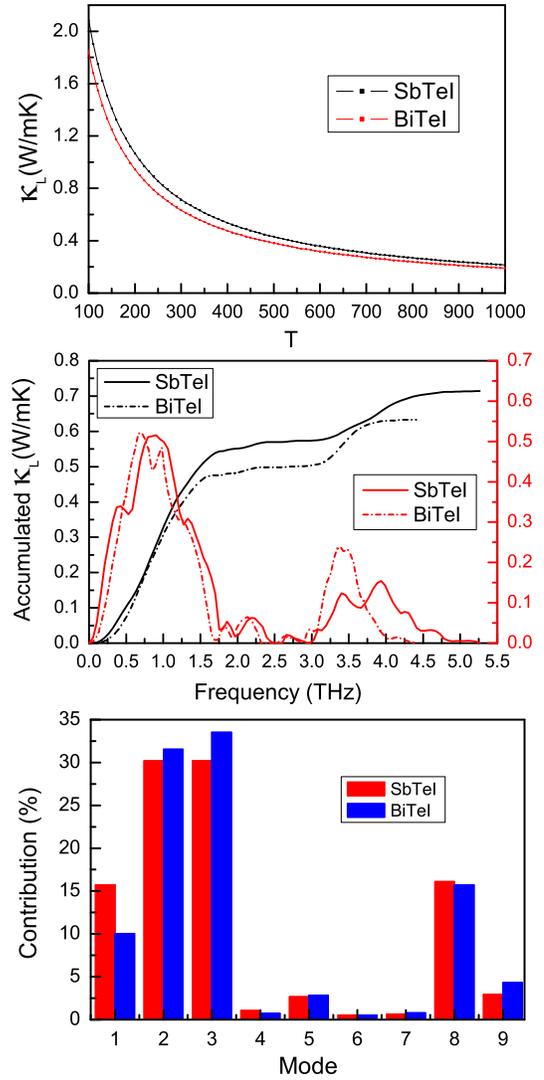}
  \caption{(Color online) Top: the lattice thermal conductivities of SbTeI  and BiTeI monolayers  as a function of temperature using GGA-PBE. Middle: the accumulated lattice thermal conductivities, and the derivatives. Bottom: phonon modes contributions toward total lattice thermal conductivity (300 K).}\label{mkl}
\end{figure}

Based on calculated energy band structures, the  electronic   transport coefficients of ATeI (A=Sb, I) monolayers  can be attained  using CSTA Boltzmann theory.  The calculated Seebeck coefficient  is independent of scattering time, while electrical conductivity depends  on  scattering time. \autoref{s} shows the Seebeck coefficient S,  electrical conductivity with respect to scattering time  $\mathrm{\sigma/\tau}$ and  power factor with respect to scattering time $\mathrm{S^2\sigma/\tau}$ as a function of doping level using both GGA and GGA+SOC at room temperature.
By simply shifting  Fermi level into conduction  or valence bands, the n- or p-type doping level can be simulated   within the framework of  rigid band approach. The approximation is effective in low doping level\cite{q30,q31,q32}.
It is found that SOC has a slightly enhanced  effect on n-type  Seebeck coefficient, but has a obviously reduced influence on p-type Seebeck coefficient.  According to \autoref{dos}, the slope of density of states (DOS) of valence bands of BiTeI near the energy band gap with GGA+SOC  decreases with respect to one with GGA,  which leads to reduced S.
Similar results can be found in SbTeI.
For $\mathrm{\sigma/\tau}$, the detrimental influence caused by SOC can be observed in p-type and low n-type doping.  A signally reduced effect on p-type power factor  can be caused by SOC, while a neglectful influence on n-type power factor can be observed.
\begin{figure*}
  \includegraphics[width=15.0cm]{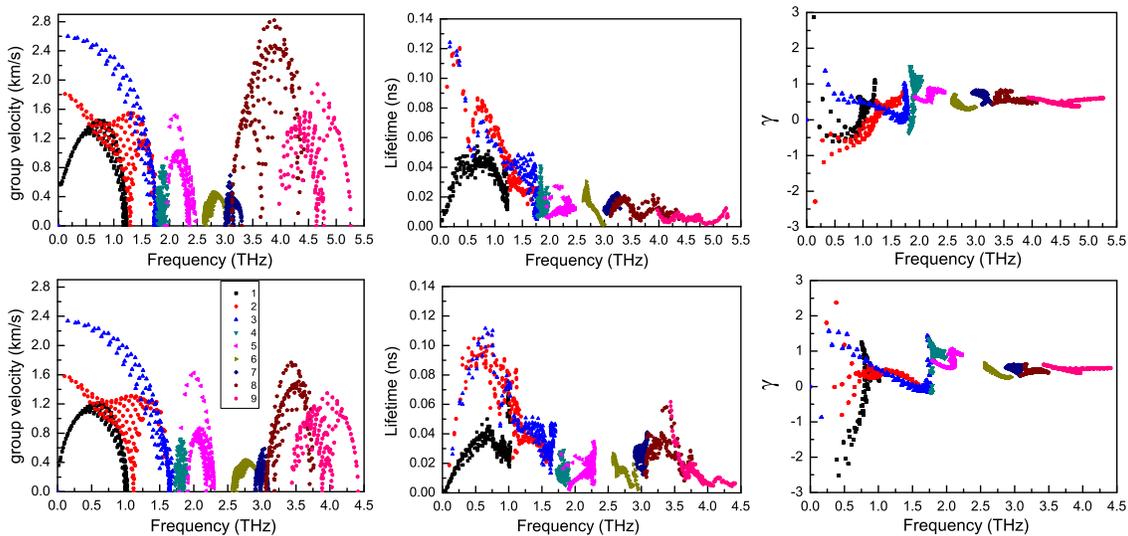}
  \caption{(Color online)From left to right, the phonon mode group velocities, phonon lifetimes (300 K)  and mode Gr$\mathrm{\ddot{u}}$neisen parameters of  SbTeI (Top) and BiTeI (Bottom) monolayers  in the first BZ.}\label{mkl1}
\end{figure*}

Based on calculated harmonic IFCs, the phonon dispersions of ATeI (A=Sb, I) monolayers, determining the group velocities and  allowed three phonon scattering processes,   are attained  along  high-symmetry pathes, which  are plotted in \autoref{phband}, together with partial DOS in \autoref{ftdos}.
The 3 acoustic and 6 optical phonon branches in the phonon spectra can be observed due to three atoms in the unit cell of  ATeI (A=Sb, I) monolayers. It is clearly seen that  the longitudinal acoustic (LA) and transverse acoustic (TA)
branches  are linear near the $\Gamma$ point,
while the z-direction acoustic (ZA) branch is quadratic. Similar results can be found in many other 2D materials\cite{q15,q17,q18,q19,q23}.
The whole  phonon  branches   move toward lower energy from SbTeI to BiTeI monolayer, which suggests the phonon dispersion becomes more localized.
The width of   acoustic branches   is 1.80 THz and 1.71 THz from SbTeI to BiTeI monolayer,  which is also maximal acoustic vibration frequency (MAVF).  The low MAVF means small  group velocities, producing low lattice thermal conductivity.
It is a very noteworthy phenomenon that the last two  optical branches have very large dispersions, which may lead to obvious
contribution to lattice thermal conductivity. According to \autoref{ftdos},
the high-frequency optical modes of ATeI (A=Sb, I) monolayers  are mainly
from A and Te vibrations, while the low-frequency optical and acoustic branches are due to the vibrations of all atoms.

The lattice thermal conductivities  of ATeI (A=Sb, I) monolayers as a function of temperature and the accumulated lattice thermal conductivity (300 K)  along with the derivatives are plotted in \autoref{mkl}. The room-temperature lattice  thermal conductivity of ATeI (A=Sb, I)  monolayers is 0.71 $\mathrm{W m^{-1} K^{-1}}$  and 0.63  $\mathrm{W m^{-1} K^{-1}}$ with the same thickness 20 $\mathrm{{\AA}}$, respectively.
To compare their  thermal conductivities with ones of  other  2D materials, the room-temperature thermal conductivity  is converted into thermal sheet conductance\cite{2dl}, and the corresponding  value  is  14.2 $\mathrm{W K^{-1}}$ and 12.6 $\mathrm{W K^{-1}}$, respectively. Their  thermal sheet conductances are lower than one of semiconducting transition-metal dichalcogenide,  group IV-VI, group-VA and group-IV (expect germanene) monolayers\cite{q23,2dl,q33}. The very low thermal sheet conductance  suggests that SbTeI  and BiTeI monolayers are potential 2D thermoelectric materials. The cumulative lattice thermal conductivity  and the derivatives show that the acoustic phonon branches dominate  lattice thermal conductivity, and the  high-frequency optical modes have obvious contribution to lattice thermal conductivity. The acoustic branches  comprise around 76.13\% for SbTeI and 75.10\% for BiTeI, respectively.
 Furthermore, the relative contributions of nine  phonon modes to the total lattice
thermal conductivity at 300K are plotted in \autoref{mkl}.
It is found that ZA branch  provides the smallest   contribution in acoustic branches, while the LA and TA branches have about the same contribution. It is a surprising thing that the last two  optical branches of SbTeI and BiTeI monolayers provide a contribution of 19.02\% and 20.03\%, which is different from usual picture with
little contribution from optical branches.  Moreover, the second optical branch has also relatively large contribution to the total lattice thermal conductivity.

According to \autoref{mkl1}, the  group velocities  of LA and TA branches  are larger than  ones of ZA branch, due to  quadratic  dispersion of ZA branch near the $\Gamma$ point. The largest  group velocity  for ZA, TA and LA branches near $\Gamma$ point is 0.58  $\mathrm{km s^{-1}}$, 1.81  $\mathrm{km s^{-1}}$ and 2.60  $\mathrm{km s^{-1}}$ for SbTeI monolayer and 0.36
$\mathrm{km s^{-1}}$, 1.58 $\mathrm{km s^{-1}}$ and 2.34 $\mathrm{km s^{-1}}$ for BiTeI monolayer. In other 2D
 materials, larger phonon group velocities near $\Gamma$ point can be found than in ATeI (A=Sb, I) monolayers, such as in blue phosphorene, arsenene, antimonene,  stanene and silicene\cite{q15,q17,q18,q19,q23}. The small phonon
group velocities can lead to lower  thermal conductivity in ATeI (A=Sb, I) monolayers than in other 2D
 materials. The  group velocities  become smaller from SbTeI to BiTeI monolayer, and then a decrescent lattice thermal conductivity can be induced.  It is very rare that optical modes have very large group velocities, especially for
 the  second and last two optical branches.   For SbTeI,  the maximum  group velocity of the fifth optical branch  is 2.82 $\mathrm{km s^{-1}}$, which is larger than maximum  group velocity of acoustic branches. These suggest that optical branches
play an important role in determining lattice thermal conductivity of ATeI (A=Sb, I) monolayers.
\begin{figure}
  \includegraphics[width=7cm]{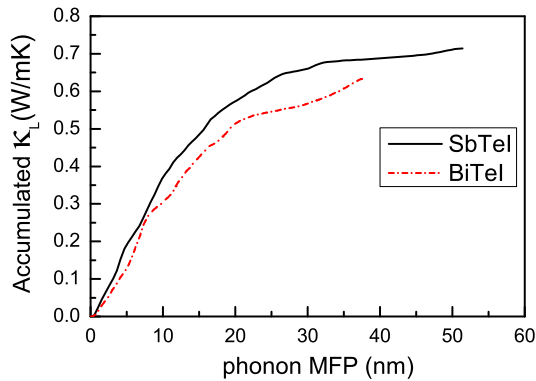}
  \caption{(Color online)Cumulative lattice thermal conductivities of ATeI (A=Sb, I) monolayers with respect to phonon mean
free path at room temperature.}\label{mfp}
 \end{figure}

Phonon lifetimes of  ATeI (A=Sb, I) monolayers at room temperature are also plotted  in \autoref{mkl1}, which  can attained by the phonon linewidth. In the single-mode relaxation time method\cite{pv4}, the phonon lifetimes and  lattice thermal conductivity  are merely proportional to each other.
The most  lifetimes of ZA branch  is shorter than ones of LA and TA branches, which leads to smaller contribution to lattice
thermal conductivity for ZA branch than LA and TA branches.  The lifetimes of most acoustic modes of SbTeI
are between 20 ps and 80 ps, and 20 ps and 100 ps for BiTeI.  The lifetimes of most optical modes are less than 20 ps for SbTeI,
but ones of high-frequency optical modes of BiTeI are well-matched with ones of  ZA branches. By considering group velocities and phonon lifetimes, smaller group velocities  for BiTeI than SbTeI lead to lower lattice
thermal conductivity.  Based on third
order anharmonic IFCs, mode Gr$\mathrm{\ddot{u}}$neisen parameters can be attained, which is shown in \autoref{mkl1}.
Most of they  are positive, especially for optical branches.  The mode Gr$\mathrm{\ddot{u}}$neisen parameters
can be used to describe  the anharmonicity of materials, and larger  $\gamma$ can lead to lower  lattice thermal conductivity because of  strong anharmonic phonon scattering. It is found that Gr$\mathrm{\ddot{u}}$neisen parameters of ZA branch of BiTeI are larger than of ones of SbTeI, which induces lower intrinsic thermal conductivity for BiTeI than SbTeI.

To quantify the contribution from phonons with various mean free
paths (MFP), the cumulative lattice thermal conductivity   with respect to phonon MFP is plotted in \autoref{mfp} at room temperature. The
cumulative lattice thermal conductivity increases  with MFP increasing, and then approaches maximum after MFP
reaches 51.4 nm for SbTeI and 37.7 nm for BiTeI, respectively.  When the characteristic length is smaller
than 51.4 nm for SbTeI and 37.7 nm  for BiTeI,  the lattice thermal conductivity can be significantly reduced. The stronger
intrinsic phonon scattering, causing  phonons to have
shorter MFP, can induce lower lattice thermal conductivity.
These critical values are very small, so scale reduction may be difficult to reduce lattice thermal conductivity.
\begin{figure}
  \includegraphics[width=7cm]{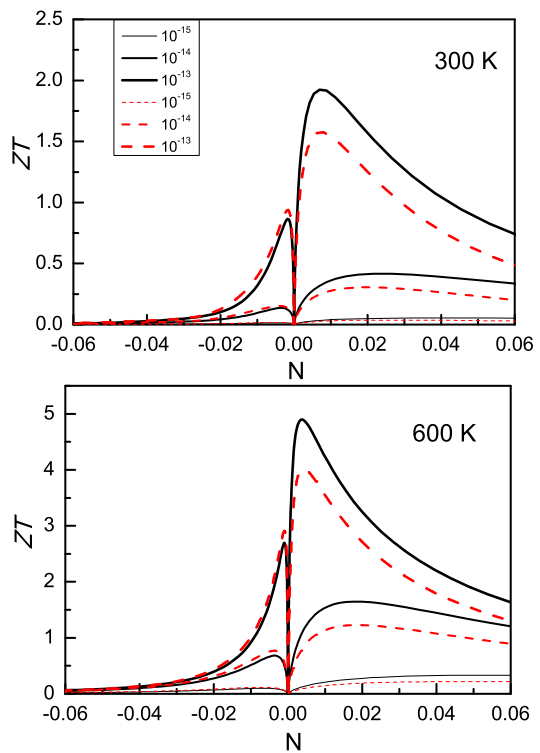}
  \caption{(Color online)At 300 and  600 K, calculated $ZT$ of SbTeI (Black lines) and BiTeI (Red lines)  monolayers  as a function of doping level using  three  scattering time  $\mathrm{\tau}$: 1 $\times$ $10^{-15}$ s, 1 $\times$ $10^{-14}$ s and 1 $\times$ $10^{-13}$ s.}\label{ft}
\end{figure}

\begin{figure}
  \includegraphics[width=7cm]{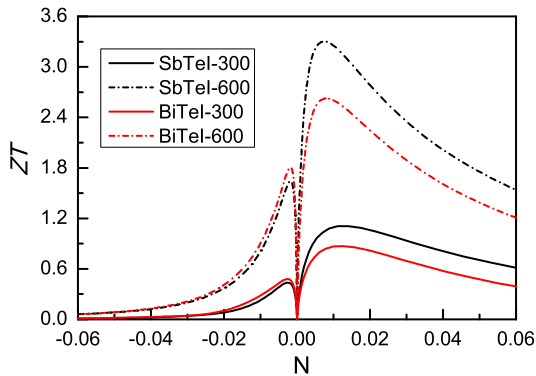}
  \caption{(Color online)At 300 and  600 K, calculated $ZT$ of ATeI (A=Sb, I) monolayers  as a function of doping level using  scattering time 3.93 $\times$ $10^{-14}$ s.}\label{ft1}
\end{figure}

To estimate the efficiency of thermoelectric conversion, the figure of merit $ZT$ is calculated, based on calculated electron and phonon transport coefficients. However,  scattering time  $\mathrm{\tau}$ is unknown, which  can be  attained by comparing experimental value  of electronic conductivity with the calculated value $\mathrm{\sigma/\tau}$.  Firstly, some empirical values, such as  1 $\times$ $10^{-15}$ s, 1 $\times$ $10^{-14}$ s  and 1 $\times$ $10^{-13}$ s, are used to calculate the power factor  and electronic thermal conductivity. Here, the electrical thermal conductivities are calculated  by the Wiedemann-Franz law with the Lorenz number of 2.4$\times$$10^{-8}$ $\mathrm{W\Omega K^{-2}}$, which is also used in bulk BiTeI\cite{q26}.  At 300 and 600 K, the possible $ZT$ of ATeI (A=Sb, I) monolayers with respect to doping level  are shown in \autoref{ft}. The $ZT$=$ZT_e$$\times$$\kappa_e/(\kappa_e+\kappa_L)$, where $ZT_e=S^2\sigma T/\kappa_e$,  which is independent of $\mathrm{\tau}$, as  an upper limit of $ZT$. The increasing $\mathrm{\tau}$ makes  $\kappa_e/(\kappa_e+\kappa_L)$ be more close to one, which leads to increasing $ZT$ with $\mathrm{\tau}$ changing from 1 $\times$ $10^{-15}$ s to 1 $\times$ $10^{-13}$ s. It is found that  the  p-type doping  for ATeI (A=Sb, I) monolayers shows more excellent  $ZT$ than  n-type doping.  Calculated results show that  SbTeI monolayer has better thermoelectric properties than BiTeI monolayer in p-type doping due to larger $ZT$, while  they show almost equivalent $ZT$ in n-type doping.
In p-type doping, a peak $ZT$  is up to 1.65 for SbTeI and 1.23 for BiTeI  using  classic $\tau$=$10^{-14}$ s at 600 K.
For bulk BiTeI, the related experimental transport coefficients can be found in ref.\cite{q34}.
Finally, the $\mathrm{\tau}$ is calculated by comparing experimental values of
the n-type conductivity of bulk BiTeI with the calculated values of $\mathrm{\sigma/\tau}$\cite{q29} at room temperature.
The scattering time is found to be  3.93 $\times$ $10^{-14}$ s, which is  also used in ATeI (A=Sb, I) monolayers.
The recomputed $ZT$ of  ATeI (A=Sb, I) monolayers   as a function of doping level with $\mathrm{\tau}$ being 3.93 $\times$ $10^{-14}$ s are plotted  in \autoref{ft1}. At room temperature,  the peak $ZT$ reaches 1.11 for SbTeI and 0.87 for BiTeI, respectively.

\begin{table}[!htb]
\centering \caption{Thermal sheet conductances of  group IV-VI,   semiconducting transition-metal dichalcogenide, group-IV, group-VA and  ATeI (A=Sb, I) monolayers. (Unit:$\mathrm{W K^{-1}}$)}\label{tab1}
  \begin{tabular*}{0.48\textwidth}{@{\extracolsep{\fill}}cccccccc}
  \hline\hline
 GeS& GeSe& SnS& SnSe&  $\mathrm{ZrS_2}$& $\mathrm{ZrSe_2}$& $\mathrm{HfS_2}$& $\mathrm{HfSe_2}$                                      \\\hline
 52.93 &  31.58&18.68&17.55&77.89 &  62.14 &  97.35&  69.38                                              \\\hline\hline
 Si& Ge& Sn&  As&Sb &Bi&  SbTeI&BiTeI
 \\\hline
 120.12&11.39&26.04&161.10&  46.62& 16.02& 14.20& 12.60
 \\\hline\hline
\end{tabular*}
\end{table}
\section{Discussions and Conclusion}
ATeI (A=Sb, I) monolayers display the Rashba effect due strong SOC, which significantly changes their conduction and valence bands. However, SOC has little effect on n-type Seebeck coefficient, leading to little influence on n-type power factor.
For bulk BiTeI, a detrimental influence on n-type Seebeck coefficient can be observed\cite{q29}. For BiTeI monolayer, the SOC not only can remove the degeneracy of conduction bands, but also can make conduction bands to be more localized. The two combined factors lead to little  influence on n-type  Seebeck coefficient caused by SOC. For both bulk and monolayer BiTeI, the reduced effect on p-type Seebeck coefficient can be produced by SOC.

Low lattice thermal conductivity is very crucial for potential thermoelectric materials. To make  a fair comparison for thermal transport capability of 2D materials,  the same thickness should be used, or the sheet thermal conductance should be adopted\cite{2dl}.
 The  sheet thermal conductances  of  ATeI (A=Sb, I) monolayers, some semiconducting transition-metal dichalcogenide, group IV-VI, group-VA and group-IV monolayers\cite{q23,2dl,q33} are summarized in \autoref{tab1}.
The sheet thermal conductances of ATeI (A=Sb, I) monolayers are  lower than that of semiconducting transition-metal dichalcogenide, group IV-VI, group-VA and group-IV monolayers except Ge monolayer, which suggests that they may  be  potential 2D thermoelectric material compared to other  well-known 2D materials.  It is found that the  small phonon
group velocities lead to  low sheet thermal conductances  of  ATeI (A=Sb, I) monolayers.

In summary, we have employed first-principles calculations and semiclassical Boltzmann transport theory to investigate the thermoelectric properties  of ATeI (A=Sb, I) monolayers. It is found that SOC can produce important effects on electronic structures and  transport coefficients  due to Rashba effect in  ATeI (A=Sb, I) monolayers.  The sheet thermal conductance is employed to make a fair comparison for  lattice thermal conductivities of different 2D materials. The sheet thermal conductances of  ATeI (A=Sb, I) monolayers are lower than one of other well-studied 2D materials due to small group velocities.
It is found that the high-frequency optical branches make a considerable contribution to the total lattice thermal conductivity.
Nanostructuring is difficult to further reduce the lattice thermal conductivity unless their characteristic lengths are less than
51.4 nm  for SbTeI and 37.7 nm for  BiTeI.  According to estimated $ZT$, ATeI (A=Sb, I) monolayers may be potential 2D thermoelectric materials, which can stimulate further experimental works to synthesize  these  2D monolayers.

\begin{acknowledgments}
This work is supported by the National Natural Science Foundation of China (Grant No. 11404391). We are grateful to the Advanced Analysis and Computation Center of CUMT for the award of CPU hours to accomplish this work.
\end{acknowledgments}

\end{document}